\documentclass[3p,times,twocolumn]{elsarticle}
 \biboptions{comma,sort&compress}
 \usepackage{amsmath}
\usepackage{graphicx}
\usepackage{ecrc}
\usepackage{braket}

\volume{00}

\firstpage{1}

\journalname{Nuclear and Particle Physics Proceedings}

\runauth{}

\jid{nppp}

\jnltitlelogo{Nuclear and Particle Physics Proceedings}

\usepackage{amssymb}

\usepackage[figuresright]{rotating}
\newcommand{\cO}{{\cal O}}

\usepackage{color}
\definecolor{orange}{RGB}{255,127,0}
\definecolor{brown}{RGB}{102,51,0}
\definecolor{myred}{RGB}{192,0,0}
\newcommand{\comment}[1]{}

\usepackage[normalem]{ulem}

\begin{document}

\begin{frontmatter}

\title{ChPT parameters from $\tau$-decay data
 $^*$}
 \cortext[cor0]{Talk given at 18th International Conference in Quantum Chromodynamics (QCD 15,  30th anniversary),  29 june - 3 july 2015, Montpellier - FR}
  \author[label1]{A. Rodr\'{i}çguez-S\'anchez\fnref{fn1}}
   \fntext[fn1]{Speaker.}
   \address[label1]{Departament de F\'{\i}sica Te\'orica, IFIC, Universitat de Val\`encia – CSIC, Apt. Correus 22085, E-46071 Val\`encia, Spain}

 \author[label2]{M. Gonz\'alez-Alonso}
\address[label2]{IPN de Lyon/CNRS, Universit\'e Lyon 1, Villeurbanne, France}

 \author[label1]{A. Pich}

\pagestyle{myheadings}
\markright{ }
\begin{abstract}
Using the updated ALEPH $V-A$ spectral function from $\tau$ decays, we determine the lowest spectral moments of the left-right correlator and extract dynamical information on order parameters of the QCD chiral symmetry breaking.
Uncertainties associated with violations of quark-hadron duality are estimated from the data, imposing all known short-distance constraints on a resonance-based parametrization.
Employing proper pinched weight functions, 
we obtain an accurate determination of the effective chiral couplings $L^{\mathrm{eff}}_{10}$ and $C^{\mathrm{eff}}_{87}$ and the dimension-six and -eight contributions in the Operator Product Expansion.

\end{abstract}

\end{frontmatter}
\section{Introduction}
Very valuable information on perturbative and non-perturbative QCD can be extracted from semileptonic $\tau$ decays \cite{Pich:2013lsa}. In this work we focus on the $V-A$ spectral function, that vanishes
identically at all orders of perturbation theory and enables us to extract non-perturbative parameters free of perturbative uncertainties.

A detailed study of the $V-A$ correlation function was given in Refs.~\cite{GonzalezAlonso:2008rf,GonzalezAlonso:2010rn,GonzalezAlonso:2010xf}, using the published ALEPH $\tau$ data \cite{Schael:2005am}.
The recent update of the ALEPH non-strange spectral function \cite{Davier:2013sfa}, which incorporates a new unfolding method that also corrects some problems in the correlations between the unfolded mass bins~\cite{Boito:2010fb}, motivates a re-analysis of the numerical estimates of low-energy effective couplings performed
in \cite{GonzalezAlonso:2008rf,GonzalezAlonso:2010rn,GonzalezAlonso:2010xf}.

The starting point is the non-strange left-right (LR) correlator \ $\Pi(q^{2})\equiv\Pi^{(0+1)}_{ud,LR}(q^{2})\equiv \Pi^{(0)}_{ud,LR}(q^{2})+\Pi^{(1)}_{ud,LR}(q^{2})$, defined as follows:
\begin{eqnarray}\nonumber
\lefteqn{
\Pi^{\mu\nu}_{ud,LR}(q) \equiv i \int d^{4}x\, e^{iqx}\bra{0}T(L^{\mu}_{ud}(x)R^{\nu \dagger}_{ud}(0))\ket{0}} &&
\\ &&\mbox{}\hskip -.8cm
= (-g^{\mu\nu}q^{2}+q^{\mu}q^{\nu})\,\Pi^{(1)}_{ud,LR}(q^{2})+q^{\mu}q^{\nu}\,
\Pi^{(0)}_{ud,LR}(q^{2}) \, ,\quad
\end{eqnarray}
where $L_{ud}^{\mu}(x)\equiv \bar{u}(x)\,\gamma^{\mu}(1-\gamma_{5})\, d(x)$ and $R_{ud}^{\mu}(x)\equiv \bar{u}(x)\,\gamma^{\mu}(1+\gamma_{5})\, d(x)$.

Using analyticity, a QCD Sum Rule \cite{Shifman:1978bx} can be used to relate the correlator in the Euclidean region, where it can be approximated by
its short distance Operator Product Expansion (OPE) \cite{Wilson:1969zs}, with its imaginary part in the Minkowskian one, accessible experimentally at low energies \cite{GonzalezAlonso:2010xf}:
\begin{eqnarray}\nonumber
\lefteqn{\int^{s_{0}}_{s_{th}}ds\, \omega(s)\,\rho(s)}
\\ \nonumber&+&\!\!\! \frac{1}{2\pi i}\,\oint_{|s|=s_{0}}ds\, \omega(s) \,\Pi^{OPE}(s)\, +\,\delta_{\mathrm{DV}}[\omega(s),s_{0}]
\\&=&\!\!
2\,f_{\pi}^{2}\,\omega(m_{\pi}^{2})\, +\,
\mathrm{Res}[\omega(s)\Pi(s),s=0]\label{sumrule}\; ,
\end{eqnarray}
%
%
where $s_{th} = 4 m_\pi^2$, 
$\omega(s)$ is an arbitrary weight function, analytic in the whole complex plane except for a possible pole at the origin, $\rho(s)\equiv\frac{1}{\pi}\operatorname{Im}\Pi(s)$ is the spectral function,  $\Pi^{OPE}(s)=\sum_{k}\frac{\mathcal{O}_{2k}}{(-s)^{k}}$ is the OPE of the correlator,
and $\delta_{\mathrm{DV}}[\omega(s),s_{0}]$ is the duality violating (DV) part, that arises from the difference between the exact correlator and its OPE representation \cite{GonzalezAlonso:2010rn,Shifman:2000jv,Cirigliano:2002jy,Cirigliano:2003kc,Cata:2005zj}:
\begin{eqnarray}\nonumber
\lefteqn{\delta_{\mathrm{DV}}[\omega(s),s_{0}]}
\\&\equiv&\frac{1}{2\pi i} \,\oint_{|s|=s_{0}}ds\, \nonumber\omega(s)\,\left[\Pi(s)-\Pi^{OPE}(s)\right]
\\&=&\;\int^{\infty}_{s_{0}}ds\;\omega(s)\,\rho(s)\, \label{eq:DV}\; .
\end{eqnarray}

The relation (\ref{sumrule}) allows us to obtain different physical parameters choosing appropriate weight functions $\omega(s)$. If we take $\omega(s)=\{s^{-2},s^{-1}\}$ the contour
integral is 0 and low-energy information can be obtained:
\begin{align}\nonumber
\int^{s_{0}}_{s_{th}}ds\, s^{-2}\rho(s)&\, =\, -\delta_{\mathrm{DV}}\left(\frac{1}{s^{2}},s_{0}\right)
+2\,\frac{f_{\pi}^{2}}{m_{\pi}^{4}}+\Pi'(0)
\\&\, =\, -\delta_{\mathrm{DV}}\left(\frac{1}{s^{2}},s_{0}\right)+ 16\, C_{87}^{\mathrm{eff}}\label{eqc87}\; ,
\\[0.5cm] \nonumber
\int^{s_{0}}_{s_{th}}ds\, s^{-1}\rho(s)&\, =\, -\delta_{\mathrm{DV}}\left(\frac{1}{s},s_{0}\right)
+\,2\frac{f_{\pi}^{2}}{m_{\pi}^{2}}+\Pi(0)
\\ &\, =\, -\delta_{\mathrm{DV}}\left(\frac{1}{s},s_{0}\right) -8\, L_{10}^{\mathrm{eff}}\; .
\label{eql10}
\end{align}
The effective couplings $L_{10}^{\mathrm{eff}}$ and $C_{87}^{\mathrm{eff}}$ are quantities that can be written in terms of low-energy $\chi$PT constants \cite{GonzalezAlonso:2008rf}. We focus
on the direct information that can be extracted aplying these sum rules with the updated ALEPH data. The determination of the corresponding $L_{10}^{r}$ and $C_{87}^{r}$ couplings at $\cO(p^6)$ is not included in this work.

Taking $\omega(s)=\{1,s\}$ we obtain:
\begin{align}
\int^{s_{0}}_{s_{th}}ds\, \rho(s)&\, =\, -\delta_{\mathrm{DV}}\left(1,s_{0}\right)+2f_{\pi}^{2}\label{wsr1}\; ,
\\[0.5cm]
\int^{s_{0}}_{s_{th}}ds\, s\,\rho(s)&\, =\, -\delta_{\mathrm{DV}}\left(s,s_{0}\right)+2f_{\pi}^{2}m_{\pi}^{2}\label{wsr2}\; ,
\end{align}
that in the $s_{0}\rightarrow\infty$ limit, where the DV parts are 0, are the first and second Weinberg Sum Rules (WSRs) \cite{Weinberg:1967kj}.

On the other hand, the contour integral is not zero anymore for $\omega(s)=\{s^{2},s^{3}\}$. It receives a contribution from the OPE of the correlator:
\begin{align}
\int^{s_{0}}_{s_{th}}ds\,s^{2} \rho(s)&\, =\,\mathcal{O}_{6}-\delta_{\mathrm{DV}}\left(s^{2},s_{0}\right)
+2f_{\pi}^{2}m_{\pi}^{4}\label{sumruleo6}\; ,
\\[0.5cm]
\int^{s_{0}}_{s_{th}}ds\,s^{3} \rho(s)&\, =\, -\mathcal{O}_ {8}-\delta_{\mathrm{DV}}\left(s^{3},s_{0}\right)
+2f_{\pi}^{2}m_{\pi}^{6}\label{sumruleo8}\; .
\end{align}
We can take advantage of it to obtain the dimension-six and -eight contributions in the OPE of the correlator.

\section{Initial estimate of the effective couplings}\label{sectionfirst} 
\begin{figure}[t]
\includegraphics[width=0.47\textwidth]{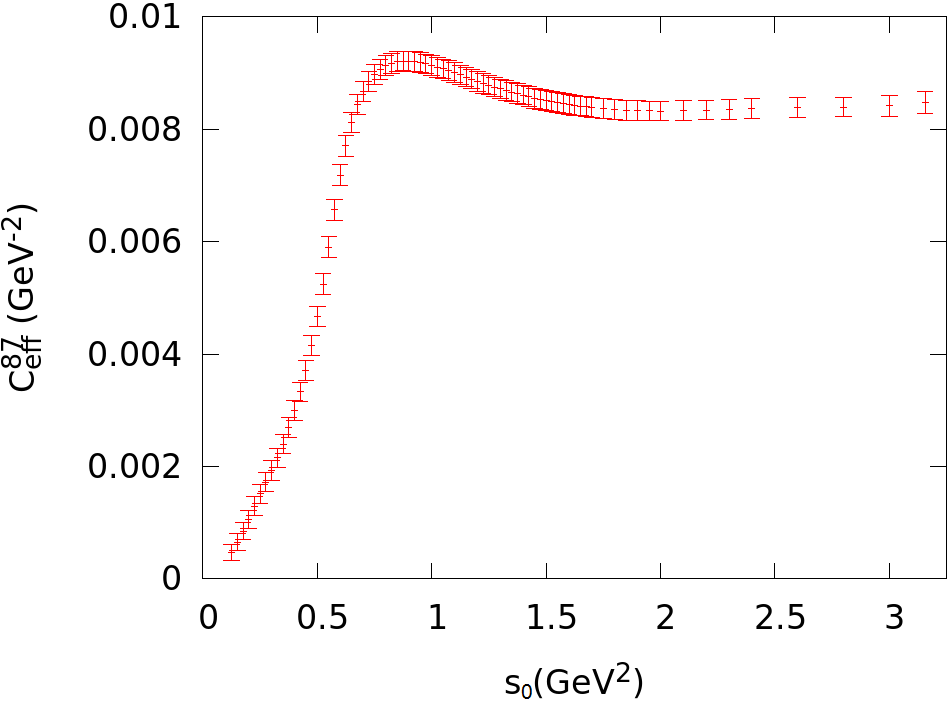}
\\[10pt]
\includegraphics[width=0.47\textwidth]{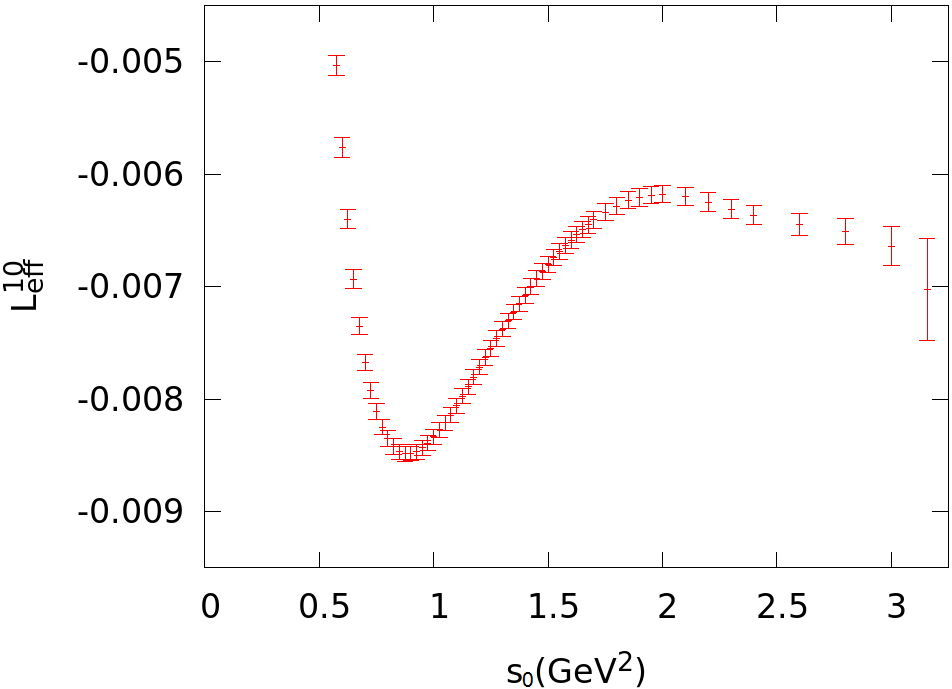}
\caption{\label{fig.c87l101}Values of $C^{\mathrm{eff}}_{87}$ and $L^{\mathrm{eff}}_{10}$ from equations (\ref{eqc87}) and (\ref{eql10}), for different $s_{0}$, neglecting duality violations.}
\end{figure}

Using the equations (\ref{eqc87}) and (\ref{eql10}), we can estimate $C^{\mathrm{eff}}_{87}$ and $L^{\mathrm{eff}}_{10}$ with the updated ALEPH spectral function \cite{Davier:2013sfa}.
If we neglect the duality violating term, 
we obtain different values of the couplings for different $s_{0}$ (Figure \ref{fig.c87l101}). As expected, 
the results are far from $s_{0}$-independent at low energies, where the DV terms are not negligible. At higher energies the curve starts to stabilise, which could indicate that duality violation effects are smaller than the experimental errors. Notice however that in the case of $L^{\mathrm{eff}}_{10}$, they are still observable.

Instead of weights of the form $s^{n}$, we can try to reduce duality violation effects using pinched weight functions \cite{LeDiberder:1992fr,Cirigliano:2003kc}, which vanish at $s=s_{0}$ (or in the vicinity), where the OPE breaks down. We will work with pinched weight functions that are linear combinations of one of the weight functions that determine an effective coupling ($s^{-2}$ or $s^{-1}$) and the weight functions
that lead to the finite WSRs
(\ref{wsr1}) and (\ref{wsr2}), which do not incorporate any new unknown physical parameters \cite{GonzalezAlonso:2008rf}:

\begin{eqnarray}
\omega_{1}(s)&\!\! =&\!\!\frac{1}{s}\;\left(1-\frac{s}{s_{0}}\right)\, ,\\
\omega_{2}(s)&\!\! =&\!\!\frac{1}{s}\;\left(1-\frac{s}{s_{0}}\right)^{2}\, ,\\
\omega_{3}(s)&\!\! =&\!\!\frac{1}{s^{2}}\;\left(1-\frac{s^{2}}{s^{2}_{0}}\right)\, ,\\
\omega_{4}(s)&\!\! =&\!\!\frac{1}{s^{2}}\;\left(1-\frac{s}{s_{0}}\right)^{2}\,
\left(1+2\,\frac{s}{s_{0}}\right)\, .
\end{eqnarray}

\begin{figure}[t]
\includegraphics[width=0.45\textwidth]{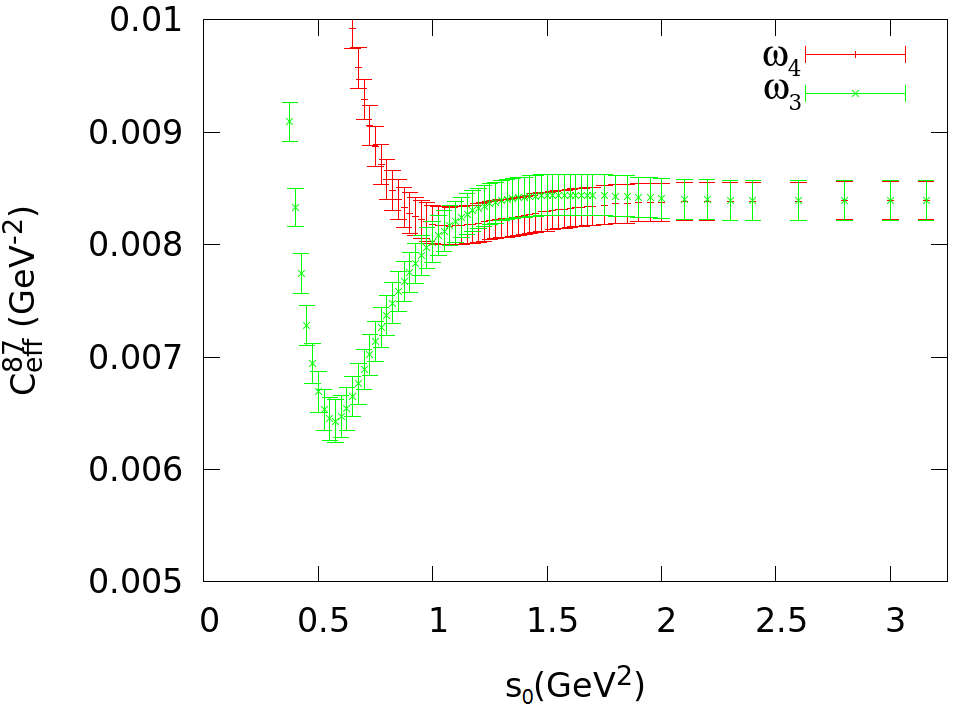}
\\[10pt]
\includegraphics[width=0.45\textwidth]{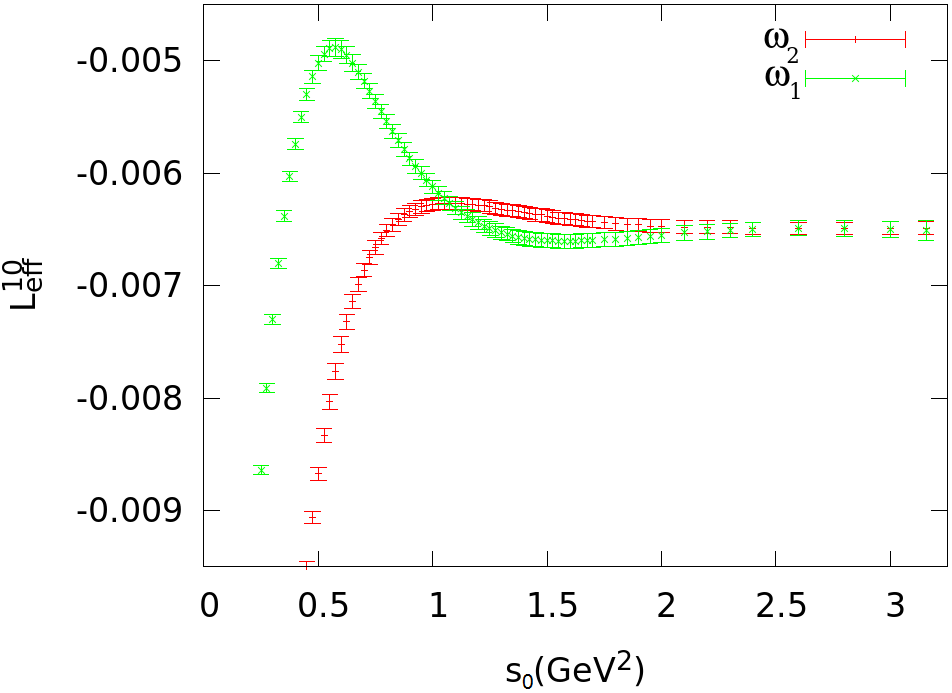}
\caption{\label{fig.c87l102}Values of $C^{\mathrm{eff}}_{87}$ and $L^{\mathrm{eff}}_{10}$, using pinched weight functions and neglecting duality violations.}
\end{figure}

We can use these pinched weight functions
to estimate the same effective couplings with reduced DV effects.
In Figure \ref{fig.c87l102} we plot the values of the couplings, for different $s_{0}$, obtained with different pinched weight functions. We observe how using them the results converge and begin to be stable
below $s=m_{\tau}^{2}$, what could indicate that duality violation effects become negligible at $s_{0} \rightarrow m_{\tau}^{2}$.
Assuming that, we obtain:

\begin{align}\label{neg1}
L^{\mathrm{eff}}_{10}\, &=\, -(6.49\pm 0.06)\,10^{-3}\; ,\\
C^{\mathrm{eff}}_{87}\, &=\, (8.39\pm 0.18)\,10^{-3} \mathrm{GeV}^{-2}\label{neg2}\; .
\end{align}

\section{Dealing with quark-hadron duality violations}\label{dualsection}

Although the stability of the determinations of $C^{\mathrm{eff}}_{87}$ and $L^{\mathrm{eff}}_{10}$ are necessary conditions for vanishing duality violations, the plateau could be temporary. We want to perform a more reliable estimate of the duality violation effects, using the last expression in Eq.~(\ref{eq:DV}).

\subsection{Parametrization}

Fortunately, although we do not have experimental access to the spectral function beyond $s=m_{\tau}^{2}$, we have some theoretical and phenomenological knowledge 
that can be used. QCD tells us that the spectral function must go to zero fast when $ s\rightarrow \infty$.
 Furthermore, we know that WSRs and the so-called Pion Sum Rule ($\pi$SR), that gives the electromagnetic mass difference of pions \cite{Das:1967it}, must be satisfied. However, this information is not enough to know the shape of the spectral function.
We need an ansatz for the spectral function compatible with that information and with the experimental data at $s\sim m_{\tau}^{2}$. The parametrization we adopt is \cite{Cata:2008ru,GonzalezAlonso:2010xf,GonzalezAlonso:2010rn}:\footnote{In references \cite{Boito:2012nt,Boito:2015fra} the s-dependence of the resonance-based model is assumed to be true for the V and A channels separately and an analysis involving 9 parameters, including the strong coupling, is performed.}
\begin{equation}\label{eqparam}
\rho(s>s_{z})\, =\,\kappa\, e^{-\gamma s}\, \sin{\left[\beta\, (s-s_{z})\right]}\; ,
\end{equation}
where $s_{z}\sim 2$ GeV$^{2}$. This parametrization incorporates the exponential fall-off and the oscillating behaviour predicted by the resonance-based model of Refs. \cite{Shifman:2000jv,Blok:1997hs,Shifman:1998rb}. Because
the exact s-dependence of the spectral function at energies $s>m_{\tau}^{2}$ is not known, we take Eq. (\ref{eqparam}) as an ansatz to estimate DV uncertainties, trying to absorb the different possible shapes that the spectral
function can have, compatible with the theoretical and experimental constraints.

Fitting the parameters given in (\ref{eqparam}) to the ALEPH data in the interval $s\in(1.7 $ GeV$^{2},m_{\tau}^{2})$ we obtain:
\begin{equation}
\chi^{2}_{min}\, =\, 8.52\sim 9\, =\,\mathrm{d.o.f.}\; ,
\end{equation}
which indicates that the parametrization is compatible with the ALEPH spectral function. In fact, the fit with the updated data is more reliable than the previous one,
where a value of $\chi^{2}_{min}/\mathrm{d.o.f.}\ll 1$ was obtained \cite{GonzalezAlonso:2010rn}.

\subsection{Selection of acceptable spectral functions}
Following the procedure described in \cite{GonzalezAlonso:2010rn}, we create $2 \cdot 10^{7}$ randomly distributed tuples of the parameters $(\kappa,\gamma,\beta,s_{z})$, in a rectangular region large enough to contain all the possible acceptable tuples, which we defined as those satisfying each of the following conditions \cite{GonzalezAlonso:2010rn}:

- The tuples must be within the 90\% C.L. region in the fit to the experimental ALEPH spectral function, in the interval $s\in(1.7 \;\mathrm{GeV}^{2},m_{\tau}^{2})$.

- The tuples must satisfy the Weinberg and the $\pi$ Sum Rules.


\section{Determination of physical parameters including DV uncertainties}

For every accepted tuple we have an acceptable spectral function\footnote{Given by the ALEPH data below $s_{z}$ and by the parametrization used above that value.} that can be used in Eqs.~(\ref{eqc87}), (\ref{eql10}), (\ref{sumruleo6}) and (\ref{sumruleo8}), with $s_{0}=s_{z}$, to obtain  different acceptable values of the physical parameters. We build histograms with those results, and from them we obtain:
\begin{align}
C^{\mathrm{eff}}_{87}&\, =\, (8.406 ^{+0.007}_{-0.006} \pm 0.18) \cdot 10^{-3}\, \mathrm{GeV}^{-2} \label{result1}
\nonumber\\&\, =\, (8.41  \pm 0.18) \cdot 10^{-3}\, \mathrm{GeV}^{-2}\; ,\\
L^{\mathrm{eff}}_{10}&\, =\, (-6.50 ^{+0.02}_{-0.03} \pm 0.08) \cdot 10^{-3}
\nonumber\\&\label{result1.5}
   \, =\,  (-6.50 ^{+0.08}_{-0.09}) \cdot 10^{-3}\; ,\\
\mathcal{O}_{6}&\, =\, (-5.4 ^{+3.0}_{-2.4} \pm 1.3) \cdot 10^{-3}\, \mathrm{GeV}^{6}\nonumber
\\&\, =\, (-5.4 ^{+3.3}_{-2.7}) \cdot 10^{-3}\, \mathrm{GeV}^{6}\; ,\\
\mathcal{O}_{8}&\, =\, (-5 ^{+8}_{-9} \pm 2) \cdot 10^{-3}\, \mathrm{GeV}^{8}
\nonumber\\&\, =\, 
      (-5^{+8}_{-9}) \cdot 10^{-3}\, \mathrm{GeV}^{8} \label{result2}\; ,
\end{align}
where the first error corresponds to DV uncertainties, computed from the dispersion of the histograms, and the second error is the experimental one.

\begin{figure}[h!]
\includegraphics[width=0.41\textwidth]{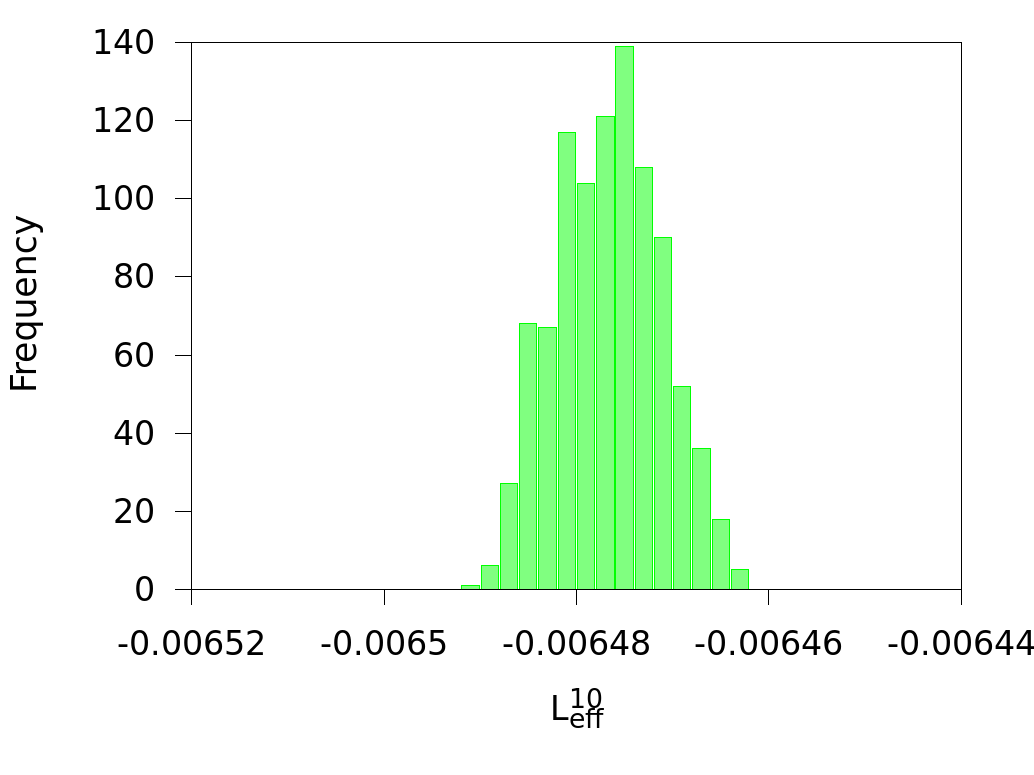}\\
\includegraphics[width=0.41\textwidth]{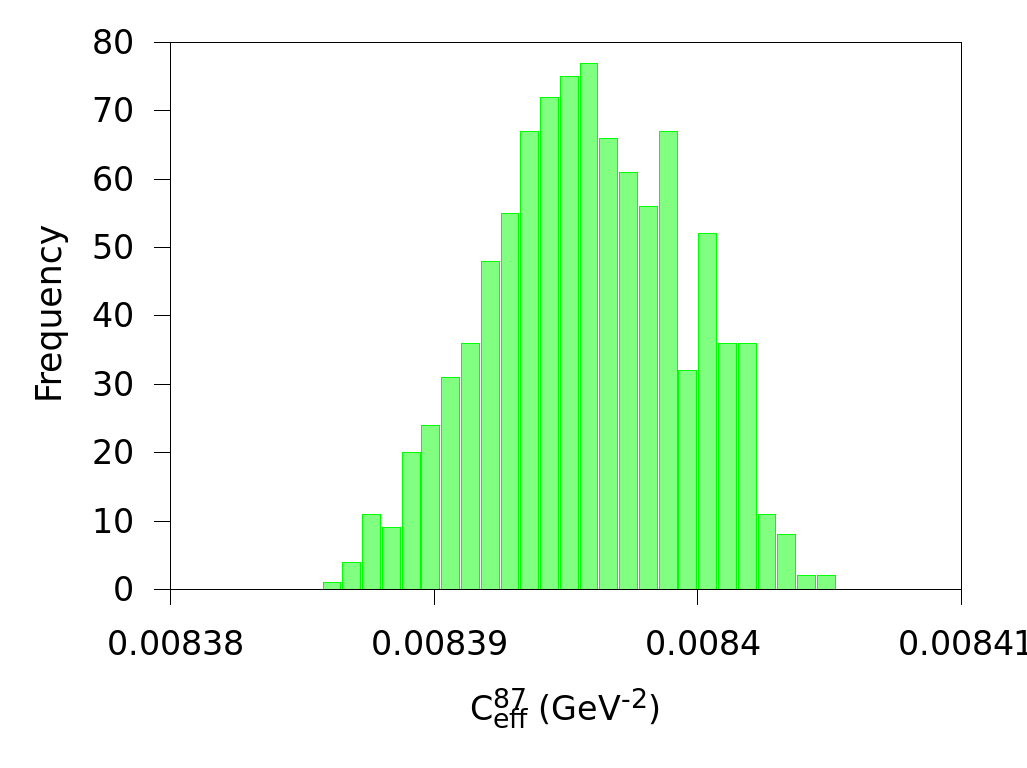}\\
\includegraphics[width=0.41\textwidth]{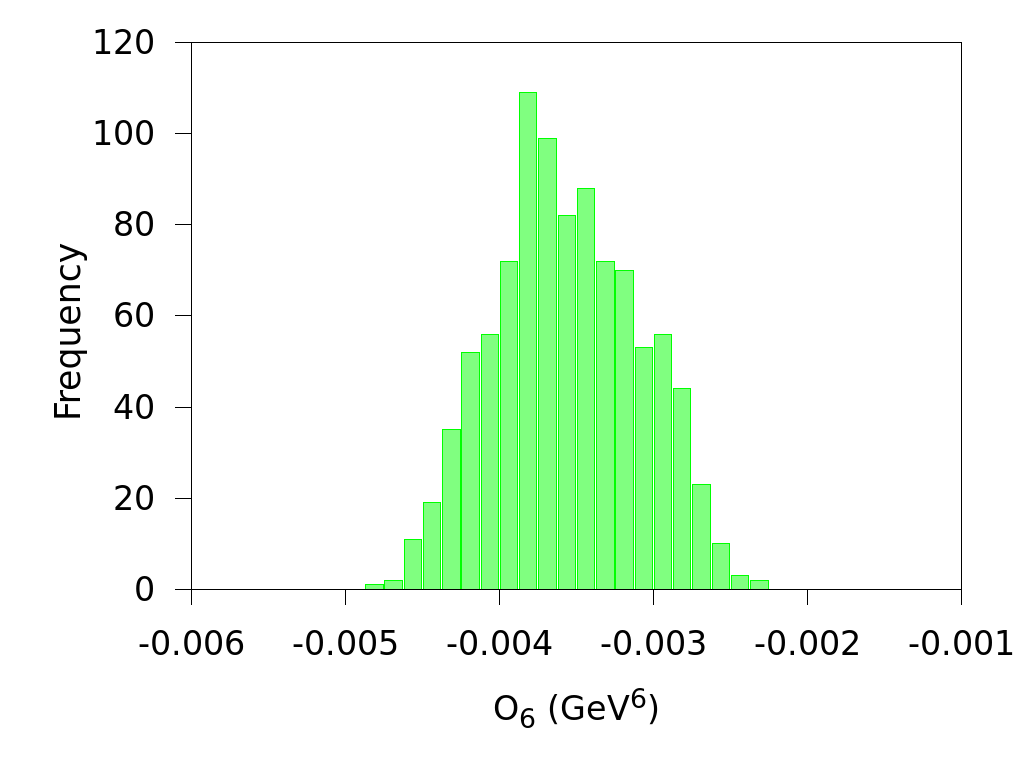}\\
\includegraphics[width=0.41\textwidth]{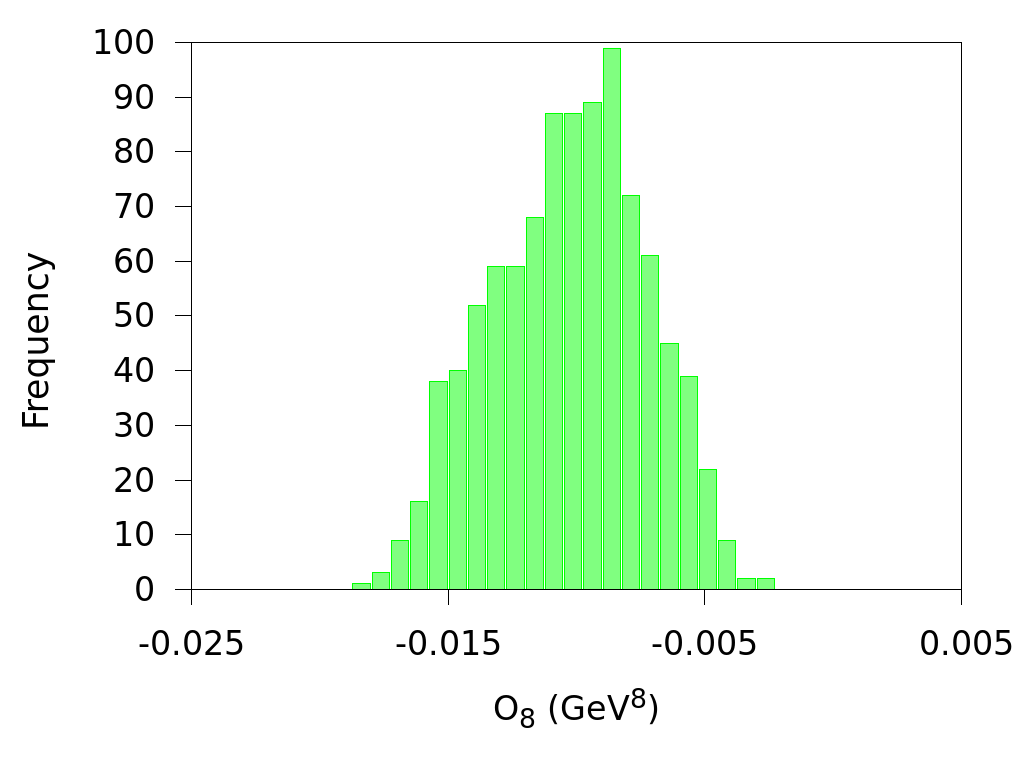}
\caption{\label{dist2}Statistical distributions of $L^{\mathrm{eff}}_{10}$, $C^{\mathrm{eff}}_{87}$, $\mathcal{O}_{6}$ y $\mathcal{O}_{8}$ for the tuples accepted using pinched weight functions.}
\end{figure}

We can reduce the DV uncertainties with the pinched weight functions used in Section \ref{sectionfirst} \cite{GonzalezAlonso:2010xf}. Following the same method with them, we obtain new distributions of acceptable physical parameters (Figure \ref{dist2}). From these new distributions we get:
\begin{align}
C^{\mathrm{eff}}_{87}&\, =\, (8.396 ^{+0.004}_{-0.004} \pm 0.18) \cdot 10^{-3}\, \mathrm{GeV}^{-2}
\nonumber\\&\label{resultpw1}
\, =\, (8.40  \pm 0.18) \cdot 10^{-3}\, \mathrm{GeV}^{-2}\; ,
\\ \nonumber
\\ &\, =\,     (-6.48 \pm 0.05) \cdot 10^{-3} \; ,
\label{resultpw1.5}\\
\mathcal{O}_{6}&\, =\, (-3.6 ^{+0.5}_{-0.5} \pm 0.5) \cdot 10^{-3}\, \mathrm{GeV}^{6}
\nonumber\\ 
&\, =\, (-3.6 \pm 0.7) \cdot 10^{-3}\, \mathrm{GeV}^{6}\; ,\label{resultpw1.9}
\\
\mathcal{O}_{8}&\, =\, (-1.0 ^{+0.3}_{-0.3} \pm 0.3) \cdot 10^{-2}\, \mathrm{GeV}^{8}
\nonumber\\
&\, =\, 
      (-1.0 \pm 0.4) \cdot 10^{-2}\, \mathrm{GeV}^{8}\label{resultpw2}\; .
\end{align}

\section{Conclusions}
We have determined $C^{\mathrm{eff}}_{87}$ and $L_{10}^{\mathrm{eff}}$ using the updated ALEPH spectral functions \cite{Davier:2013sfa} with the methods developed in \cite{GonzalezAlonso:2008rf,GonzalezAlonso:2010rn,GonzalezAlonso:2010xf}. Our preliminary results, obtained using pinched weight functions in a
statistical analysis that includes possible duality violation uncertainties compatible with the ansatz of Eq.~\eqref{eqparam}, are
(\ref{resultpw1}) and (\ref{resultpw1.5}):
\begin{align}
C^{\mathrm{eff}}_{87}&=(8.40  \pm 0.18) \cdot 10^{-3}\, \mathrm{GeV}^{-2}\; ,\\
L^{\mathrm{eff}}_{10}&=(-6.48 \pm 0.05) \cdot 10^{-3}\; .
\end{align}
We find that DV errors are indeed subdominant, and thus we find a good agreement with the estimates given in Eqs. (\ref{neg1}), (\ref{neg2}), where DV was simply neglected, or Eqs. (\ref{result1}), (\ref{result1.5}), where pinched weight functions were not used. Furthermore, the results are in agreement with those obtained in Ref.~\cite{Boito:2015fra} using the same experimental data but a different analysis of DV effects (see footnote~2):
\begin{align}
C_{87}^{\mathrm{eff}}&\, =\, (8.38  \pm 0.18) \cdot 10^{-3}\, \mathrm{GeV}^{-2}\; , \\
L_{10}^{\mathrm{eff}}&\, =\,  (-6.45 \pm 0.05) \cdot 10^{-3}\; ,
\end{align}
and the ones obtained with the non-updated spectral function in \cite{GonzalezAlonso:2010xf}:
\begin{align}
C_{87}^{\mathrm{eff}}&\, =\, (8.17  \pm 0.12) \cdot 10^{-3}\, \mathrm{GeV}^{-2}\; , \\
L_{10}^{\mathrm{eff}}&\, =\,  (-6.44 \pm 0.05) \cdot 10^{-3}\; .
\end{align}
We notice nonetheless that the error in $C_{87}^{\mathrm{eff}}$ was significantly smaller when the old dataset was used.

The statistical analysis used allows a determination of the dimension-six and -eight OPE contributions. We show here again our preliminary results, Eqs. (\ref{resultpw1.9}) and (\ref{resultpw2}):
\begin{align}
\mathcal{O}_{6}&\, =\, 
			(-3.6 \pm 0.7) \cdot 10^{-3}\, \mathrm{GeV}^{6}\; ,\\
\mathcal{O}_{8}
 &	\, =\, (-1.0 \pm 0.4) \cdot 10^{-2}\, \mathrm{GeV}^{8}\; ,
\end{align}
also compatible with the determinations performed in Refs.~\cite{Boito:2015fra,GonzalezAlonso:2010xf}.

\section*{Acknowledgments}
This work has been supported in part by the Spanish Government and ERDF funds from
the EU Commission [Grants No.  FPA2011-23778, FPA2014-53631-C2-1-P], by the Spanish
Centro de Excelencia Severo Ochoa
Programme [Grant SEV-2014-0398] and by Generalitat
Valenciana under Grant No.  PROMETEOII/2013/007. M.G.-A. is grateful to the LABEX Lyon Institute of Origins (ANR-10-LABX-0066) of the Universit\'e de Lyon for its financial support within the program ANR-11-IDEX- 0007 of the French government.
\clearpage

\end{document}